# Visualizing the Effect of an Electrostatic Gate with Angle-Resolved Photoemission Spectroscopy

Frédéric Joucken[1]†, Jose Avila[2]†, Zhehao Ge[1], Eberth Quezada[1], Hemian Yi[2], Romaric Le Goff[3], Emmanuel Baudin[3], John L. Davenport[1], Kenji Watanabe[4], Takashi Taniguchi[4], Maria Carmen Asensio[5*], Jairo Velasco Jr.[1*]

[1] Department of Physics, University of California, Santa Cruz, California 95060, USA

[2] ANTARES Beamline, Synchrotron SOLEIL & Université Paris-Saclay, L'Orme des Merisiers, Saint Aubin-BP 48, 91192 Gif-sur-Yvette Cedex, France

[3] Laboratoire Pierre Aigrain, Département de physique de l'ENS, Ecole normale supérieure, PSL Research University, Université Paris Diderot, Sorbonne Paris Cité, Sorbonne Universités, UPMC Univ. Paris 06, CNRS, 75005, Paris, France

[4] Advanced Materials Laboratory, National Institute for Materials Science, 1-1 Namiki, Tsukuba, 305-0044, Japan

[5] Instituto de Ciencia de Materiales de Madrid, Sor Juana Inés de la Cruz, 328049 Madrid, Spain

**Abstract:**

Electrostatic gating is pervasive in materials science, yet its effects on the electronic band structure of materials has never been revealed directly by angle-resolved photoemission spectroscopy (ARPES), the technique of choice to non-invasively probe the electronic band structure of a material. By means of a state-of-the-art ARPES setup with sub-micron spatial resolution, we have investigated a heterostructure composed of Bernal-stacked bilayer graphene (BLG) on hexagonal boron nitride and deposited on a graphite flake. By voltage biasing the latter, the electric field effect is directly visualized on the valence band as well as on the carbon 1s core level of BLG. The band gap opening of BLG submitted to a transverse electric field is discussed and the importance of *intra*layer screening is put forward. Our results pave the way for new studies that will use momentum-resolved electronic structure information to gain insight on the physics of materials submitted to the electric field effect.

**Main Text:**

In the realm of two-dimensional (2D) materials and beyond, the electric field effect is ubiquitous.[1–3] All modern electronics use this effect in transistors to modulate the conductivity of a channel by aligning its electronic bands with those of the source and the drain. Although studied for a long time,[4] graphene became a watershed when its ambipolar behavior was demonstrated by electrostatic gating,[3] triggering the intense development of 2D materials research.

Angle-resolved photoemission spectroscopy (ARPES) is the standard method to probe the electronic band structure of materials.[5] The photoelectric effect and the conservation of the photoelectrons' momentum parallel to the probed surface enables direct visualization of the electronic bands. ARPES is particularly adapted to 2D materials, since the absence of dispersion in the z-direction makes the interpretation of the spectra straightforward. Until now however, the modulation of the electronic structure of a material by field effect has never been imaged directly *via* ARPES. This is firstly because before the advance of 2D materials, gated 2D electron gases only occurred at buried interfaces, inaccessible to photoemission.[2,6,7] Once 2D electron gases were brought to the surface,[3] two major technological challenges remained. The first was to put them on a substrate that allows gating and is sufficiently flat for ARPES measurements. This was solved with the advent of hexagonal boron nitride (*h*BN) as a supporting substrate.[8,9] The second challenge was to enhance the spatial resolution of ARPES to permit probing of micron-scale devices made with 2D materials.

In this letter, we account for ARPES measurements of the electronic band structure of a material whose Fermi level is modulated by electric field effect. The nanoARPES setup, depicted in Fig. 1A, allows ARPES imaging of the sample with submicron resolution as well as high-resolution ARPES measurements at desired locations.[10,11] The schematic of our device, a heterostructure composed of Bernal-stacked bilayer graphene (BLG) on top of an insulating $h$BN crystal and a graphite flake that acts as a back-gate, is depicted in Fig. 1B. The device components from this schematic are visible in the nanoARPES spatial image shown in Fig. 1C.[10] This image illustrates that nanoARPES can be used to locate various parts of a micron-scale device unambiguously. Notably, the graphite back-gate plays an important role in enabling the visualization of the field effect with ARPES. First, the proximity of the graphite back-gate permits large charge carrier densities ($n$) because of the strong capacitive coupling between the BLG and graphite flake. Secondly, we believe that the graphite back-gate suppresses photo-induced doping,[12,13] which would cancel gate-induced doping and preclude any visualization of field effect *via* ARPES, as indicated by the recent work of Campos et al.[14] We performed source-drain resistance measurements as a function of the back-gate voltage $V_G$ (Fig. 1D) *in situ* (in the same ultrahigh vacuum chamber as the one used for the nanoARPES measurements). Clear bipolar resistance behavior is seen, thus demonstrating electric field effect *via* the graphite back-gate.

Our main results are presented in Fig. 2. ARPES E-k spectra obtained with $V_G$ set to 0 V, +12 V and -10 V, are displayed in Figs. 2A, 2B, and 2C, respectively. The direction of the momentum space that is probed is indicated in the inset of Fig. 2A. The Fermi level in each of the spectrum is denoted by a dashed black line. Each spectrum displays a noticeable asymmetry in momentum around K≃1.70 Å$^{-1}$. For Fig. 2A ($V_G$=0 V), we observe that the charge neutrality point (CNP) coincides with the Fermi level, implying an undoped sample. In contrast, for Figs. 2B and 2C ($V_G$=+12 V and $V_G$=-10V, respectively), a shift of the electronic bands with respect to the Fermi level is seen. A positive gate voltage shifts the Fermi level into the conduction band, thus *n*-doping the BLG sheets (Fig. 2B) while a negative gate voltage shifts the Fermi level into the valence band, thus *p*-doping the BLG sheets (Fig. 2C). Quantitatively, we find that the energy differences

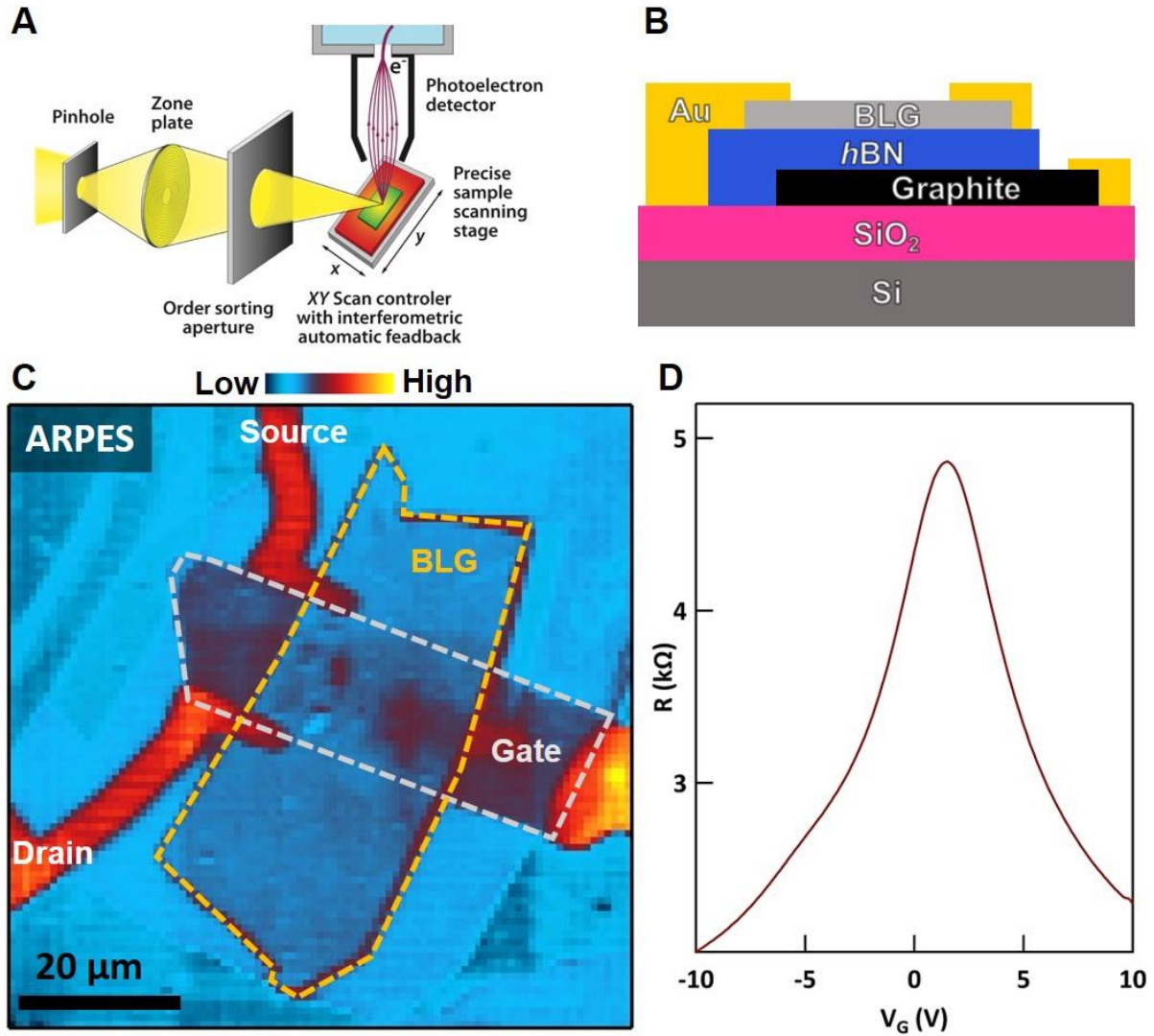

Figure 1 **Measurement scheme, device schematic, and electrical characterization**. (**A**) Schematic of angle-resolved photoemission spectroscopy setup with submicron spatial resolution (nanoARPES). The beam is focused by a zone plate, followed by an order sorting aperture. The sample sits on a XY stage with interferometric feedback, allowing nanoARPES spatial imaging.[10] (**B**) Device schematic for our Bernal-stacked bilayer graphene (BLG)/hexagonal boron nitride (*h*BN)/graphite heterostructure resting upon a $SiO_2$/Si substrate. The underlying graphite flake functions as a back-gate ($V_G$). (**C**) NanoARPES spatial image of the device on which the various components are indicated. Data acquired with a photon energy of 100 eV and kinetic energy was integrated on a 11 eV window.[10] (**D**) R-$V_G$ curve obtained on the device *in situ*.

between the Fermi level and the CNP ($E_F$-$E_{CNP}$) are, respectively, 0.165+/-0.035 eV and -0.130+/-

0.045 eV for Figs. 2B and 2C. We also show in Figs. 2D, 2E, and 2F the experimental Fermi

surface measurements obtained at the same gate voltages as the spectra in Figs. 2A-C (0V, +12V, and -10V). We observe on these Fermi surfaces a clear shift in the spectral weight across the K point (indicated by a vertical dashed black line) as the gate voltage is modulated.

Additional insights on our above observations can be gained by superimposing calculated tight-binding (TB) electronic bands (dashed blue line) on the experimental spectra (Figs. 2A, 2B, and 2C).[10] Since no observable difference in the shape of the bands is measured between the three cases, the same TB parameters are used, and the bands are simply shifted vertically by the corresponding values measured for the CNP shifts. We find excellent agreement between the measured spectra and the calculated bands. The asymmetry in momentum around K that veils the shifts of the bands can be attributed to photoemission matrix element effects, unavoidable with our measurement configuration.[15,16] More precisely, for binding energies below the CNP, the matrix element darkens the branch of the bands corresponding to $k_x>K$, whereas for binding energies above the CNP, the branch is darkened for $k_x<K$. Remarkably, because of this effect, our Fermi surface measurements unambiguously demonstrate the shift of the bands and doping polarity change due to gate modulation. Indeed, for the *n*-doped case (Fig. 2E) the measured spectral intensity is greater for $k_x>K$ because the Fermi level is higher than the CNP. In contrast, the intensity is greater for $k_x<K$ in the *p*-doped case because the Fermi level is lower than the CNP.

To further investigate gate modulation of the electronic states within our BLG device we measured the C1s core level binding energy *via* X-ray photoemission spectroscopy (XPS). XPS spectra obtained under different gate voltages and with a micron-sized X-ray spot focused on the BLG are shown in Fig. 3A.[10] The data reveal a clear dependence upon gate modulation: application of a positive (negative) gate voltage shifts the spectra to higher (lower) binding energy. The variation in binding energy of the BLG C1s level as a function of gate voltage ($V_G$=0 V being the reference) is plotted in Fig. 3B (black dots), together with measured CNP shifts of the valence bands (red triangles). As expected,[17] the core level and the valence band shifts follow the same trend, indicating a rigid shift of all the electronic states upon gate modulation. The dependence of the energy shifts on *n* (upper axis in Fig. 3B) follows the theoretical shift (grey curve in Fig. 3B) obtained from a parallel plate capacitor model.[10] Finally, we note that the width of the BLG C1s XPS peak changes slightly with the gate voltage. We believe this is due to local heating induced by the exposure to the x-ray beam.[10]

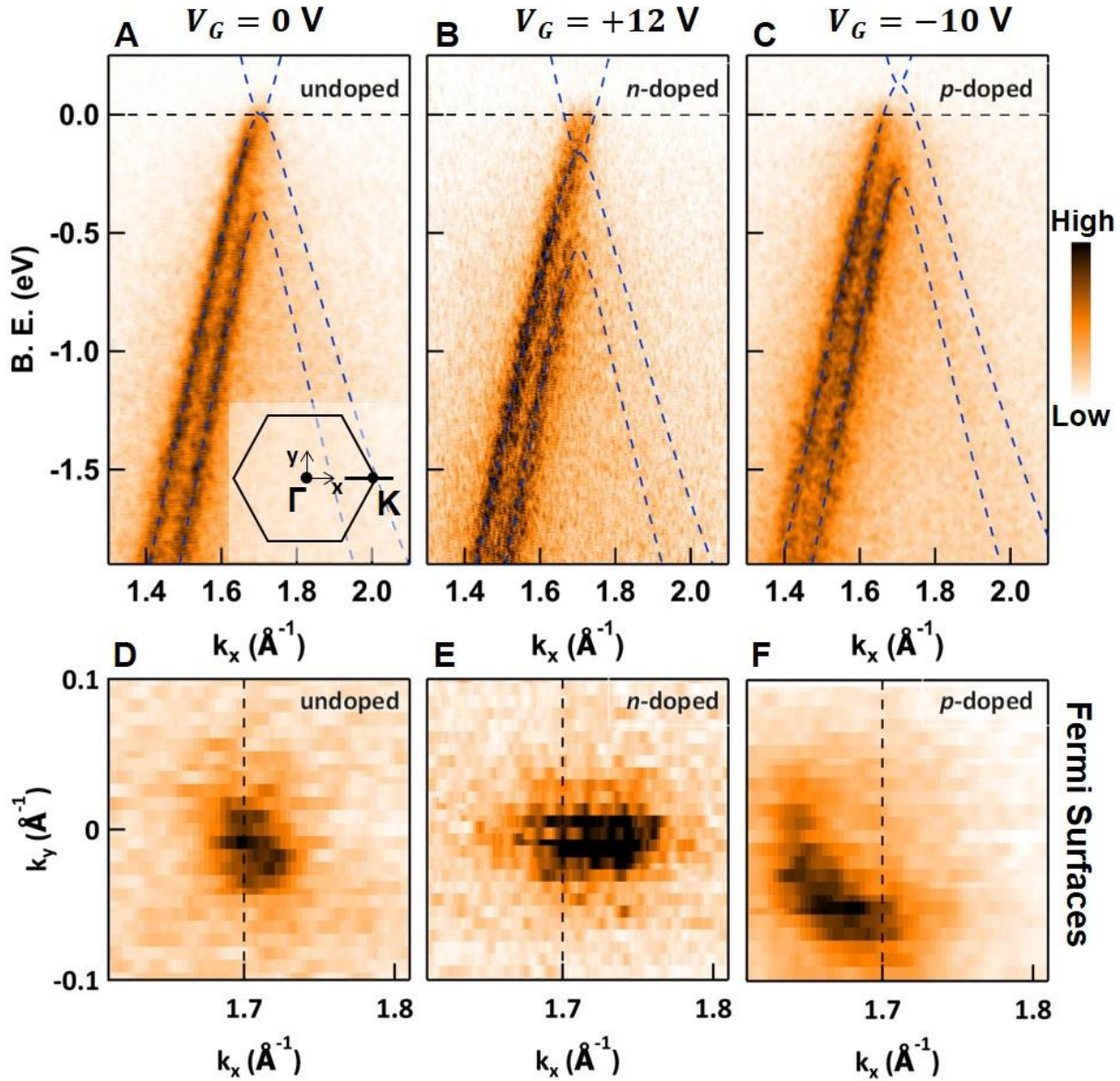

Figure 2: **Gate-induced shifting of BLG bands observed with nanoARPES**. (**A**)-(**C**) ARPES E-k spectra obtained at gate voltages of 0 V (undoped), +12V (*n*-doped), and -10 V (*p*-doped), respectively. Left axis is the binding energy (B. E.); bottom axis is the in-plane momentum. The Fermi level (B. E.=0 eV) is denoted by a dashed black line in each spectrum. The tight-binding bands[10] are superimposed on each spectrum and are offset in energy by 0.0 eV, +0.165 eV and -0.130 eV, respectively. Inset of (**A**) indicates the direction of the measurement. (**D**)-(**F**) Experimental Fermi surfaces for the same gate voltages as in (**A**)-(**C**). The dotted black lines indicate the position of the K point. A shift of the spectral weight across the K point is observed when going from *n*-doped to *p*-doped. Data acquired with a photon energy of 100 eV.

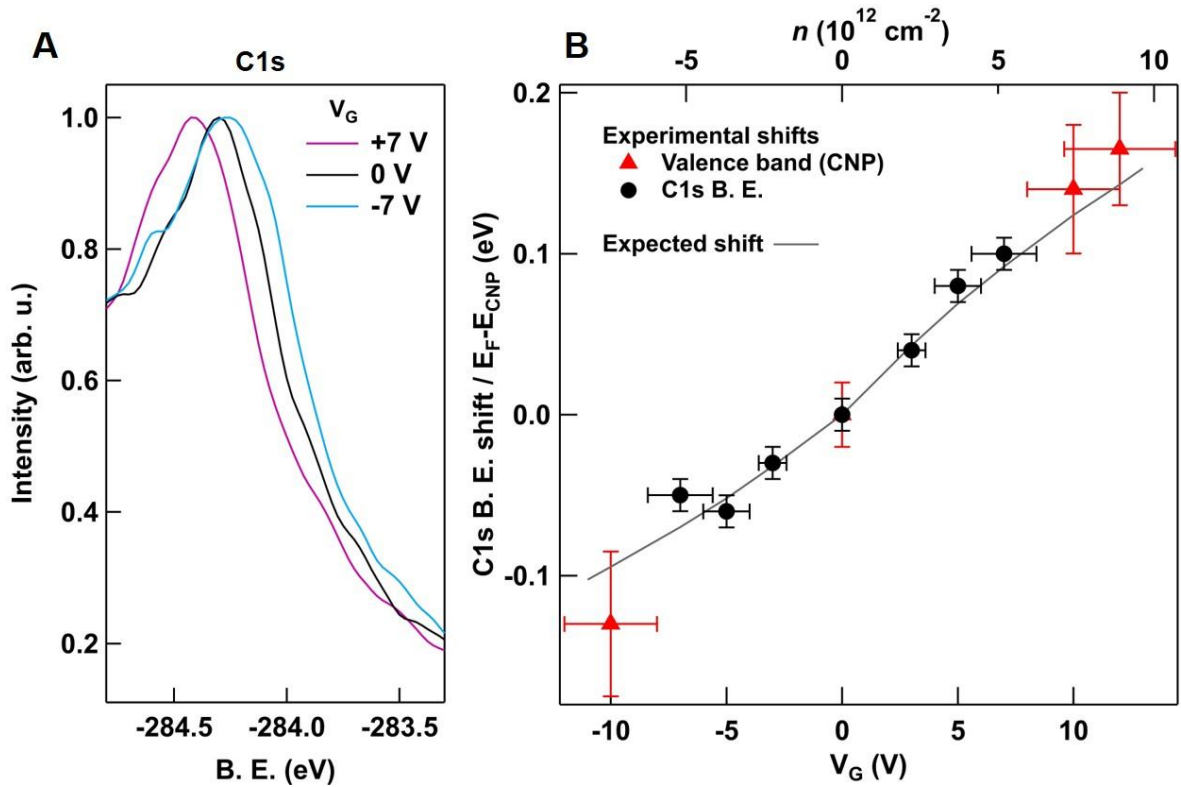

Figure 3: **Binding energy shift of core level and valence band states upon electrostatic gating.** (**A**) C1s X-ray photoemission spectroscopy (XPS) spectra of BLG under various gate voltages. Data acquired with a photon energy of 350 eV. (**B**) Summary of the experimental and expected shifts. Black dots denote C1s binding energy shifts; red dots represent shifts of the charge neutrality point (CNP) measured by nanoARPES. The grey curve is the computed expected shift for this device, assuming a parallel plate capacitor model.[10]

We now discuss the band gap formation in BLG and its apparent absence in our measurements. The possibility of opening a gap in BLG by submitting it to a transverse electric field was first established theoretically by McCann.[18] The gap ($U_g$) is produced by the potential difference between the two layers U ($U_g = U\gamma_1/\sqrt{U^2 + \gamma_1^2}$) that is induced by the electric field that breaks the interlayer symmetry.[19] Early ARPES work on BLG grown on SiC has shown the presence and modulation of a gap upon molecular doping [20,21], which coarsely tunes U and *n* in an invasive manner. Subsequently, evidence for an electric field-induced gap was reported for single-gated devices (similar to the gate configuration of our sample) *via* infrared spectroscopy measurements. However, because single-gated samples cannot realize a large electric field in conjunction with low *n*, the optical selection rules preclude direct measurement of the gap in these experiments.[22–24] A significant advance in the tuning and measurement of this gap was achieved by optical experiments that used double-gated samples. This gate configuration enabled independent control of electric field and $n$[25,26] and the results were successfully explained by self-consistent TB approaches[24,27] that take screening into account.

Inspired by these prior works, we reproduce in Fig. 4A TB (orange dashed line)[18] and GW-corrected density-functional theory (DFT-GW, red solid line)[28] calculations for $U_g$ as a function of *n,* that consider the effect of screening. To understand how these models compare to our experiment we show in Fig. 4B an ARPES E-k spectrum obtained at $V_G$=+12 V ($n \simeq 9 \times 10^{12}$ cm$^{-2}$). Evidently, we do not observe a gap. Given our experimental resolution and the apparent broadening of the measured BLG bands, we estimate the upper bound on the band gap to be ~100 meV. For $n \simeq 9 \times 10^{12}$ cm$^{-2}$, the TB and the DFT-GW calculations predict $U_g$ = 146 meV and 81 meV, respectively (vertical dashed line). We superimposed TB bands with the corresponding values for U (156 meV and 81 meV, respectively) on the ARPES E-k spectrum of Fig. 4B, for comparison. The absence of the band gap and the upper bound (~100 meV) in our measurements allows us to reject the TB predictions for the band gap ($U_g$ = 146 meV). The lack of agreement with the TB prediction suggests that the *intra*layer polarization, which is considered in the DFT-GW calculation, is essential for predicting the gap value measured by experiment.

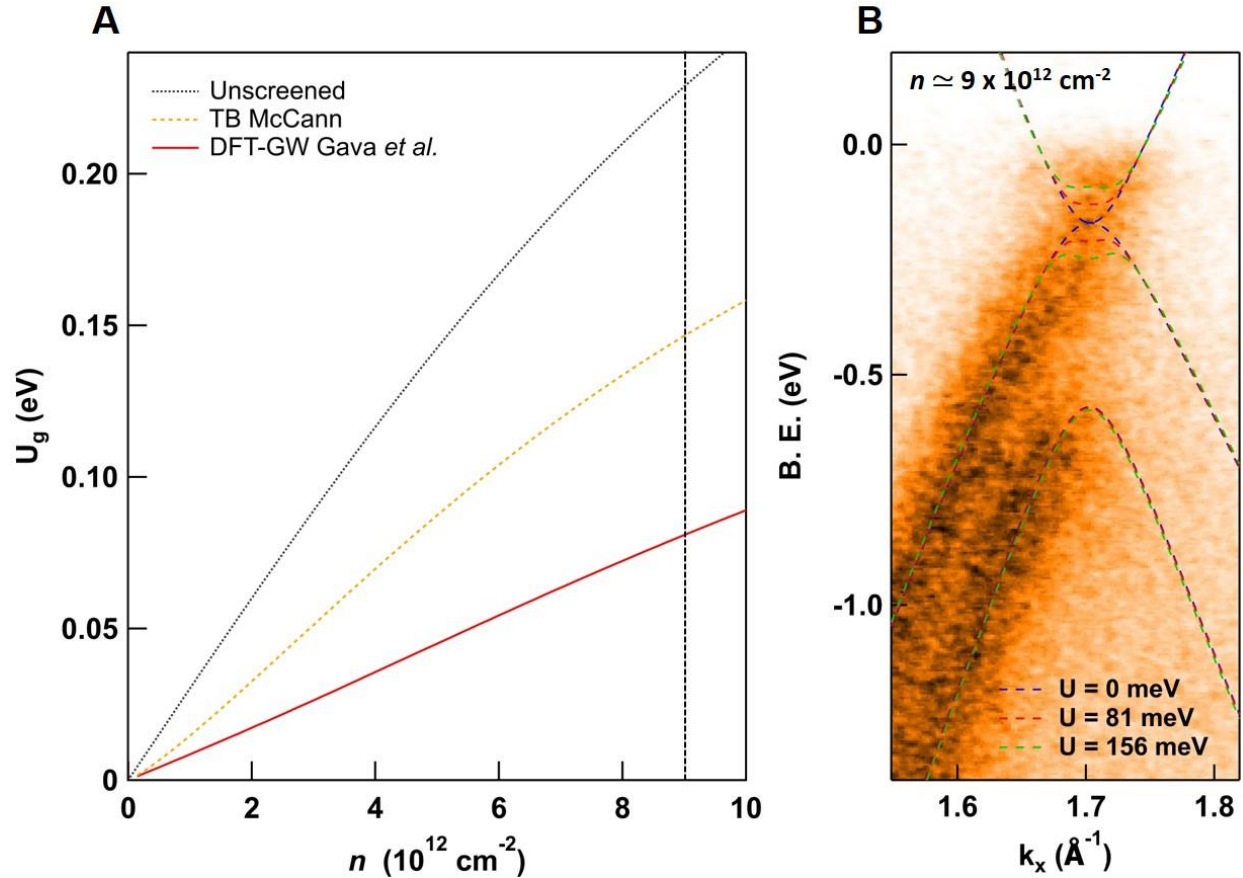

Figure 4: **Comparison between computed band gap in gated BLG with measured E-k spectrum.** (**A**) Computed band gap ($U_g = U\gamma_1/\sqrt{U^2 + \gamma_1^2}$) as a function of charge carrier density $n$ using self-consistent TB (orange dashed line)[18] and DFT-GW (solid red line)[28] methods. The calculation for the unscreened band gap (dotted grey line) is also shown. (**B**) Experimental ARPES E-k spectrum obtained at $V_G$=+12 V (data acquired with a photon energy of 100 eV), corresponding to $n \simeq 9 \times 10^{12}$ cm$^{-2}$. This $n$ is denoted as a vertical dashed line in (**A**). The TB bands for the TB- and DFT-predicted values of U (156 meV and 81 meV, respectively) are superimposed on the experimental spectrum. The lack of agreement between the TB bands and the experimental spectrum allows us to reject the value for the gap predicted by the TB model.

Given our experimental data (Fig. 4B), we cannot however exclude a bandgap smaller than predicted by the DFT-GW calculations. Such a reduction in the band gap would arise from the influence of the substrate, structural defects, or polymer residue on the sample's surface. The TB parameter describing the intralayer asymmetry between A and B sites in each graphene layer ($\delta_{AB}$), could be different from zero because of the influence of the *h*BN substrate.[19,29,30] Specifically, for a nonzero U, a finite $\delta_{AB}$ reduces the band gap amplitude.[29] It is also plausible that, although our beam spot size is small (~500 nm), we are integrating over spatial regions that contain AB-BA domain walls. These structural defects have been shown[31] to host topologically protected edge states that spoil the electric field-induced gap.[32,33] Additionally, it has been shown that the density of these domain walls can be significant even in exfoliated samples.[34] Thus, we cannot exclude the presence of AB-BA domain walls in our sample. Finally, the presence of polymer residue on our sample might also influence the formation of the bandgap, as recently observed in molecularly-doped bilayer graphene.[35] Given the small residual doping in our sample determined with electronic transport measurements ($< 2 \times 10^{12}$ cm$^{-2}$)[10] we expect this effect to be small ($<$ 20 meV).[35]

Our results demonstrate the unprecedented combination of ARPES and electric field effect. Ambipolar behavior commonly observed in electronic transport measurements was revealed by direct imaging of the electronic bands in momentum space. This allowed direct investigation of the band gap formation in BLG submitted to a transverse electric field. The significance of intralayer screening for understanding this band gap was underscored. The combination of ARPES and electric field effect on 2D materials permits the vast community of researchers studying devices composed of these materials direct access to their modulated electronic structure in momentum space.

**Associated content:**

Supporting information
    Sample preparation
    NanoARPES and XPS measurements
    Tight-binding fitting
    Determination of error bars
    Determination of the theoretical energy shifts
    Electron transport characterization of BLG/hBN/graphite heterostructure
    XPS imaging additional data
    Bilayer graphene C1s XPS peak width variation

**Corresponding authors:** *Correspondence to: mc.asensio@gmail.com, jvelasc5@ucsc.edu

**Author Contributions:** F.J., J.A., and M.C.A. conceived the experiment. Z.G., E.Q., and F.J. fabricated the sample, under J.V.J's supervision. J.A., F.J., and H.Y. performed the nanoARPES/XPS experiments, under M.C.A.'s supervision. R.L.G. and E.B. performed the *in situ* transport experiment. J.L.D. performed the AFM measurements. K.W. and T.T. provided the *h*BN crystals. F.J. and J.A. analyzed the data. F.J. and J.V.J wrote the manuscript. All authors discussed the paper and commented on the manuscript. †F.J. and J.A. contributed equally.

**Funding Sources:** R.L.G and E.B. acknowledge the ANR-14-CE08-018-05 funding "GoBN". J.V.J acknowledges support from UCOP. Atomic Force Microscope images were taken with an instrument acquired from Contract W911NF-17-1-0473 from the Army Research Office. BLG/hBN/graphite stacks were assembled in a glove box that was acquired from Contract W911NF-17-1-0473 from the Army Research Office.

**Acknowledgments:** We thank F. Zhang for helpful discussions and A. K. M. Newaz for technical support in the sample's preparation.

**Competing interests:** Authors declare no competing interest

**TOC Graphic:**

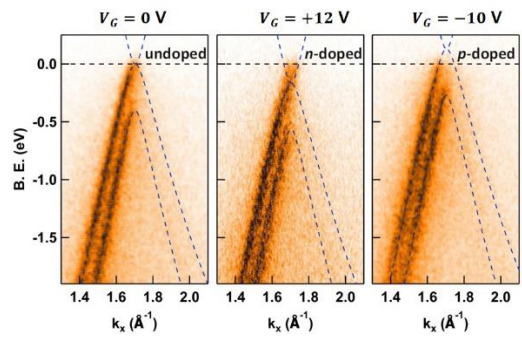